\documentclass[conference]{IEEEtran}
\IEEEoverridecommandlockouts

\usepackage{amsmath}
\usepackage{amssymb}
\usepackage{amsfonts}
\usepackage{graphicx}
\usepackage{float}
\usepackage[table, dvipsnames]{xcolor}
\usepackage{adjustbox}
\usepackage{booktabs}
\usepackage[short]{optidef}
\usepackage{multirow}
\usepackage{graphicx}
\usepackage{algorithm}
\usepackage{algpseudocode}
\usepackage{caption}
\usepackage{subcaption}
\usepackage{tabularray}
\usepackage{arydshln}
\usepackage{tikz}
\usepackage{multirow}
\usepackage{makecell}
\usepackage{booktabs}
\usepackage{color,soul}

\IEEEoverridecommandlockouts\IEEEpubid{\makebox[\columnwidth]{979-8-3503-7786-6/24/\$31.00~\copyright~2024~IEEE } \hspace{\columnsep}\makebox[\columnwidth]{ }}
        
\begin{document}

\title{Efficient Intrusion Detection: Combining $\chi^2$ Feature Selection with CNN-BiLSTM on the UNSW-NB15 Dataset}

\author{
\IEEEauthorblockN{
Mohammed Jouhari\IEEEauthorrefmark{1},
Hafsa Benaddi\IEEEauthorrefmark{1},
Khalil Ibrahimi\IEEEauthorrefmark{2},
}
\IEEEauthorblockA{
\IEEEauthorrefmark{1}LSIA Laboratory, Moroccan School of Engineering Sciences (EMSI), Tanger, Morocco\\
\IEEEauthorrefmark{2}Ibn Tofail University, Faculty of Sciences, Laboratory of Research in Informatics (LaRI), Kenitra, Morocco\\
}
E-mails: m.jouhari@emsi.ma, h.benaddi@emsi.ma, ibrahimi.khalil@uit.ac.ma
}

\maketitle

\begin{abstract}
 Intrusion Detection Systems (IDSs) have played a significant role in the detection and prevention of cyber-attacks in traditional computing systems. It is not surprising that this technology is now being applied to secure Internet of Things (IoT) networks against cyber threats. However, the limited computational resources available on IoT devices pose a challenge for deploying conventional computing-based IDSs. IDSs designed for IoT environments must demonstrate high classification performance, and utilize low-complexity models. Developing intrusion detection models in the field of IoT has seen significant advancements. However, achieving a balance between high classification performance and reduced complexity remains a challenging endeavor. In this research, we present an effective IDS model that addresses this issue by combining a lightweight Convolutional Neural Network (CNN) with bidirectional Long Short-Term Memory (BiLSTM). Additionally, we employ feature selection techniques to minimize the number of features inputted into the model, thereby reducing its complexity. This approach renders the proposed model highly suitable for resource-constrained IoT devices, ensuring it meets their computation capability requirements. Creating a model that meets the demands of IoT devices and attains enhanced precision is a challenging task. However, our suggested model outperforms previous works in the literature by attaining a remarkable accuracy rate of 97.90\% within a prediction time of 1.1 seconds for binary classification. Furthermore, it achieves an accuracy rate of 97.09\% within a prediction time of 2.10 seconds for multiclassification.
\end{abstract}

\IEEEpeerreviewmaketitle
\section{Introduction}
The Internet of Things (IoT) refers to the rapidly expanding network of interconnected devices capable of collecting, exchanging, and processing data with minimal human intervention. These devices range from simple sensors and actuators to complex computing devices integrated into everyday objects, including home appliances, wearable technology, and industrial machinery. The significance of IoT lies in its ability to enable smarter decision-making, automate tasks, and enhance the quality of life and efficiency of operations across various sectors such as healthcare, agriculture, manufacturing, and smart cities \cite{10122600}.

While the proliferation of IoT devices offers immense potential, it also poses significant security challenges. Intrusion Detection Systems (IDS) are essential for detecting unauthorized access, misuse, or compromise of networked IoT devices \cite{s19081935}. IDS serves as a monitoring mechanism that analyzes network traffic or system behaviors, flagging anomalies that could indicate a security breach. The primary role of IDS in the IoT ecosystem is to ensure the integrity, confidentiality, and availability of data and services, which are critical for maintaining trust in IoT applications.

However, implementing effective IDS in the IoT landscape is particularly challenging due to the resource-constrained nature of many IoT devices \cite{10279198}. These devices often have limited processing power, memory, and energy resources, which constrain the capability to deploy traditional, computationally-intensive IDS solutions. Additionally, the heterogeneity of IoT devices, the massive scale of the networks, and the need for real-time detection further complicate the establishment of robust intrusion detection mechanisms.

Machine learning (ML) has emerged as a powerful tool to enhance the capabilities of IDS in such a constrained environment by leveraging intelligent algorithms that can learn from data patterns and identify anomalies with minimal manual intervention  \cite{9685598, jouhari:hal-04432030}. Nevertheless, tailoring ML solutions that are lightweight and efficient enough to operate on resource-constrained IoT devices remains an active and urgent area of research. This paper aims to contribute to the ongoing efforts to enhance the security of IoT ecosystems by focusing on the development of an IDS framework specifically optimized for resource-constrained IoT devices using machine learning techniques. We present our contributions in this study as follows: 
\begin{itemize}
    \item Firstly, we propose an innovative hybrid approach that combines lightweight CNNs and bidirectional LSTM to achieve efficient and accurate intrusion detection in IoT networks. 
    \item Secondly, we develop a customized CNN architecture with reduced layers and parameters, enabling efficient inference on IoT devices.  
    \item Thirdly, we address the issue of imbalanced data in the UNSW-NB15 dataset by utilizing a weighted loss function, thereby enhancing the performance of our multiclassification model. The weight assigned to each attack is determined based on its proportion in the entire dataset. 
    \item Additionally, we propose the utilization of Chi square feature selection to reduce the number of features inputted into the model, thereby reducing the complexity of our model. 
    \item The reduction in the number of features leads to a decrease in the prediction time of our model during the inference phase, making it more suitable for resource-constrained IoT devices.
\end{itemize}
The subsequent parts of this paper are organized as follows: In Section II, the most significant works in the literature relevant to our research are discussed. Section III offers an elaborate explanation of our suggested hybrid intrusion detection method, which encompasses the lightweight CNN-BiLSTM architecture and the feature selection algorithm under consideration. It also outlines the dataset utilized and specifies the evaluation metric. Section IV showcases the experimental outcomes and analysis of our discoveries. Lastly, Section V concludes the paper by summarizing our findings and highlighting our commitment to advancing the concepts introduced in this paper through future research endeavors.

\section{Related work}
In the field of intrusion detection systems (IDS) tailored for resource-constrained IoT devices, the combination of Convolutional Neural Networks (CNN) and Bidirectional Long Short-Term Memory networks (BiLSTM) has been explored to leverage the respective capabilities of feature extraction and sequence learning. Such a hybrid approach is particularly suitable for processing the complex and voluminous data generated in IoT environments, where traditional IDS methods may be infeasible due to limited computational resources.

Authors in \cite{Omarov2023} proposed a CNN-BiLSTM hybrid model explicitly designed for network anomaly detection in IoT, employing the UNSW-NB15 dataset to evaluate its performance. This study exemplifies the push towards more sophisticated machine learning architectures that can handle the intricacies of network traffic data while still being efficient enough for deployment on devices with constrained resources. While the specific adoption of CNN-BiLSTM models for IoT devices using the UNSW-NB15 dataset is limited, related studies have introduced novel approaches that aim to achieve the same goal of developing lightweight IDS for IoT systems. For example, Authors in \cite{10180926} proposed a novel technique merging incremental principal component analysis (IPCA) with Self Adjusting Memory KNN (SAM-KNN) to create an intrusion detection model that is both effective and suitable for resource-limited environments.

Additionally, approaches like an online incremental support vector data description (OI-SVDD) and an adaptive sequential extreme learning machine (AS-ELM) have emerged as alternative solutions for crafting a lightweight NIDS catered to industrial IoT devices, demonstrating the research community's commitment to tackling the challenges of IoT security with innovative, resource-aware techniques \cite{9768133}. Moreover, promising strides continue to broaden the scope of machine learning-based IDS for IoT, aiming to enhance network security by utilizing efficient ML-enabled systems. The study by Al-Ambusaidi et al. on an effective ML-IDS reflects the ongoing endeavor to secure IoT networks and applications with machine learning methodologies, although it does not focus specifically on the CNN-BiLSTM model and UNSW-NB15 dataset \cite{10.1007/s00500-023-09452-7}.

\section{Proposed method}
\subsection{Description of the UNSW-NB15 dataset}
The UNSW-NB15 dataset provides an authentic portrayal of contemporary network traffic, encompassing a diverse range of activities. These activities span regular operational behaviors as well as nine distinct modern attack types, including Denial-of-Service (DoS), backdoors, and reconnaissance. Notably, the dataset’s well-structured organization comprises ten distinct data categories, ensuring a balanced representation of both normal and malicious traffic. As a pivotal resource, it plays a vital role in advancing and evaluating intrusion detection systems. For a comprehensive understanding of each data type, readers are encouraged to refer to Table \ref{tab:unsw_categories}.

\begin{table}[b!]
\centering
\renewcommand{\arraystretch}{1.2}
\caption{Categories of the UNSW-NB15 Dataset}
\label{tab:unsw_categories}
\begin{tabular}{clp{5.3cm}}
\toprule
\textbf{No.} & \textbf{Type} & \textbf{Description} \\ 
\bottomrule
1 & Normal & Routine, legitimate network traffic \\ \hline
2 & Analysis & Attempts to uncover vulnerabilities through web application attacks \\ \hline
3 & Backdoor & Covertly bypasses authentication for unauthorized remote access \\ \hline
4 & DoS & Disrupts system resources through memory-based attacks \\ \hline
5 & Exploits & Leverages network errors or bugs for malicious purposes \\ \hline
6 & Fuzzers & Crashes systems by overwhelming them with random data \\ \hline
7 & Generic & Exploits block-cipher vulnerabilities using hash functions \\ \hline
8 & Reconnaissance & Gathers information to evade security controls \\ \hline
9 & Shellcode & Executes code to exploit software vulnerabilities \\ \hline
10 & Worms & Self-replicating malware that spreads across systems \\ \bottomrule
\end{tabular}
\end{table}

In our study, we use the UNSW-NB15 dataset, which contains 257,673 data instances split into 175,341 training samples and 82,332 testing samples. Both sets share the same attack types, as shown in Table \ref{tab:unsw_categories}. Each data raw has 44 features, including core network packet information (features \#1-\#30), additional insights (features \#31-\#42), and labeled features (\#43-\#44) for supervised learning. This dataset is valuable for exploring intrusion detection in modern networks, including IoT networks in smart cities.

\begin{figure*}[t]
    \centering
    \includegraphics[width=0.92\linewidth]{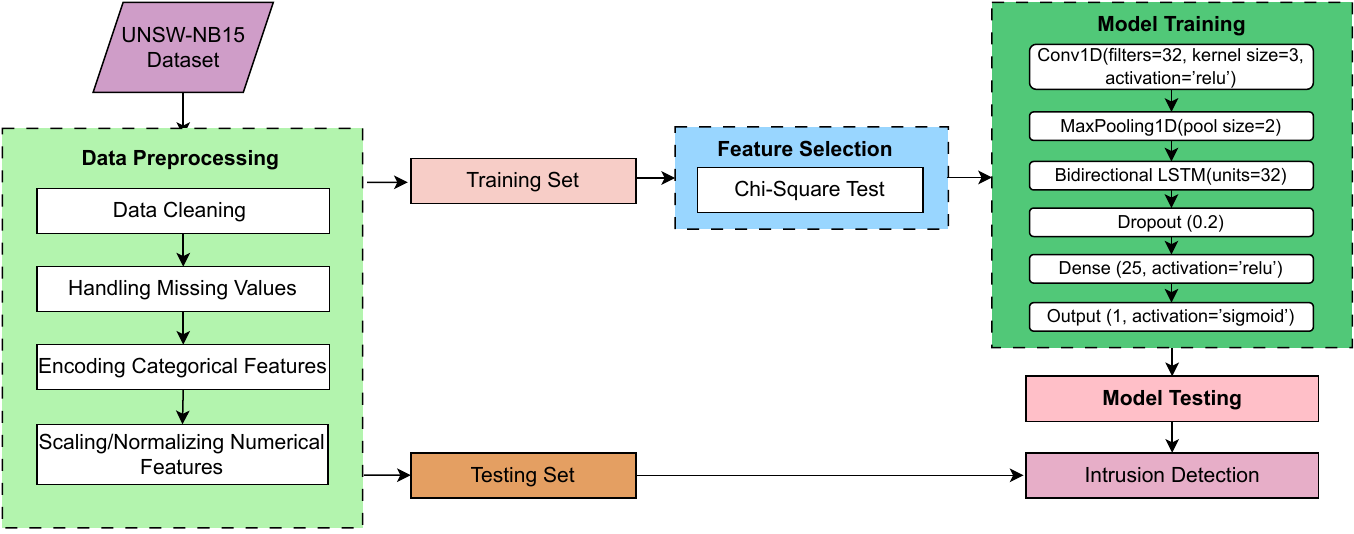}
    \caption{Proposed Model Architecture}
    \label{fig:Archi}
    \vspace{-6mm}
\end{figure*}

\subsection{Lightweight CNN-BiLSTM}
Traditional IDS often rely on rule-based systems or shallow machine learning models. However, these approaches can struggle to keep pace with the evolving nature of cyberattacks, especially in resource-constrained environments such as the Internet of Things (IoT). To address these challenges, we propose a novel hybrid model that combines the strengths of lightweight Convolutional Neural Networks (CNNs) with bidirectional Long Short-Term Memory (BiLSTM) layers \cite{tatsunami2022sequencer}. Our approach aims to enhance intrusion detection accuracy while minimizing computational complexity and memory usage. In the first stage of our model, we employ a lightweight CNN architecture specifically designed for efficient processing on edge devices. Unlike standard CNNs, this variant employs fewer layers and parameters, reducing the computational burden. The CNN’s convolutional layers extract spatial features from network traffic data, identifying patterns and relationships between different data points.

In our proposed architecture Figure \ref{fig:Archi}, the extracted features undergo a critical transformation: they are fed into a Bidirectional Long Short-Term Memory (BiLSTM) layer. This specialized layer excels at capturing temporal dependencies within sequential data, a crucial capability for network traffic analysis. Cyberattacks, often characterized by specific patterns evolving over time, necessitate such nuanced modeling. The architecture integrates both Convolutional Neural Networks (CNNs) and BiLSTM layers. The CNN initially extracts spatial features from the network traffic data, while the subsequent BiLSTM delves into the temporal dynamics. By analyzing the sequence of features bidirectionally, the BiLSTM learns the underlying patterns and identifies anomalies that deviate from normal network behavior.

The outputs from the CNN and BiLSTM layers are thoughtfully combined, leveraging the complementary strengths of both approaches. This fusion of features is then directed to a final decision-making layer, often a sigmoid classifier, which discerns the class of the network traffic (normal or malicious). The BiLSTM model comprises a pair of Long Short-Term Memory (LSTM) units. One LSTM unit processes the input sequence in the forward direction, while the other processes it in reverse. The forward LSTM analyzes the input sequence from start to end, while the backward LSTM operates conversely. At each time step, the hidden states of these two LSTMs are combined or concatenated, yielding the final hidden state sequence. This combined sequence is subsequently propagated to the subsequent layer. The mathematical formalism underlying this intricate process is elegantly expressed through a set of equations, further enriching our understanding of the CNN-BiLSTM architecture:

\begin{align}
\mathbf{h}_t^{(f)} &= \text{LSTM}^{(f)}\left(\mathbf{x}_t, \mathbf{h}_{t-1}^{(f)}, \mathbf{c}_{t-1}^{(f)}\right) \\
\mathbf{h}_t^{(b)} &= \text{LSTM}^{(b)}\left(\mathbf{x}_t, \mathbf{h}_{t+1}^{(b)}, \mathbf{c}_{t+1}^{(b)}\right) \\
\mathbf{h}_t &= [\mathbf{h}_t^{(f)}; \mathbf{h}_t^{(b)}]
\end{align}
where $\mathbf{h}_t^{(f)}$ and $\mathbf{h}_t^{(b)}$ represent the forward and backward hidden states at time step t, respectively. $\mathbf{c}_{t-1}^{(f)}$ and $\mathbf{c}_{t-1}^{(b)}$ represent the forward and backward cell states at time step t, respectively. $\mathbf{x}_t$ represents the input vector at time step t. $\mathbf{h}_t$ represented the concatenation of the forward and backward hidden states into a single vector.

\subsection{$\chi^2$ features selection}
The chi-square feature selection method is employed to identify the most informative features from the UNSW-NB15 dataset reducing the dimensionality of the input data while preserving the critical information required for accurate intrusion detection. The selected features are then used to train a CNN-BiLSTM model which combines the strengths of convolutional neural networks and bidirectional long short-term memory to effectively capture both spatial and temporal patterns in the network traffic data. The proposed approach aims to strike a balance between model accuracy and computational efficiency making it suitable for deployment on resource-constrained IoT devices. By leveraging chi-square feature selection the model can achieve high detection rates while minimizing the computational and memory requirements enabling real-time intrusion detection in IoT environments. The implementation of the chi-square feature selection was executed using the SelectKBest class from the scikit-learn library. This class offers a versatile toolset of statistical tests that can be applied to select a predetermined number of features, thereby optimizing the feature selection process for improved model performance and efficiency. Table \ref{tab:score} shows the feature scores in the UNSW-NB15 dataset using the Chi-Square feature selection method. In the training phase, we utilized the top 20 features to feed our model with data. This efficient selection significantly reduced the prediction time during the inference phase as we can see in the next section.

\begin{table}[!t]
\centering
\vspace{3mm}
\renewcommand{\arraystretch}{1.12}
\caption{Feature Scores in the UNSW-NB15 Dataset Using the Chi-Square Feature Selection Method}
\label{tab:score}
\begin{tabular}{lllll}
\toprule
\textbf{Feature} & \textbf{Score} & \textbf{Feature} & \textbf{Score} \\
\bottomrule
swin & 1.72e6 & rate & 2.48e4 \\
dwin & 1.42e6 & dpkts & 2.06e4 \\
sttl & 1.19e6 & sload & 1.75e4 \\
dload & 1.22e5 & djit & 1.16e4 \\
stcpb & 1.18e5 & sinpkt & 9.97e3 \\
dtcpb & 1.18e5 & spkts & 7.81e3 \\
dttl & 1.14e5 & ct\_state\_ttl & 6.95e3 \\
ct\_src\_dport\_ltm & 1.10e5 & response\_body\_len & 5.26e3 \\
ct\_dst\_sport\_ltm & 1.05e5 & dinpkt & 5.08e3 \\
ct\_dst\_src\_ltm & 9.99e4 & sbytes & 3.28e3 \\
ct\_srv\_dst & 8.26e4 & is\_sm\_ips\_ports & 1.12e3 \\
ct\_srv\_src & 7.89e4 & tcprtt & 3.15e2 \\
dbytes & 6.14e4 & synack & 2.55e2 \\
dloss & 5.24e4 & ct\_flw\_http\_mthd & 2.29e2 \\
ct\_src\_ltm & 5.17e4 & trans\_depth & 2.28e2 \\
ct\_dst\_ltm & 4.83e4 & ackdat & 1.10e2 \\
dmean & 4.27e4 & smean & 9.68e1 \\
sloss & 3.40e4 & ct\_ftp\_cmd & 2.47e1 \\
sjit & 2.48e4 & is\_ftp\_login & 2.17e1 \\
\bottomrule
\end{tabular}
\vspace{-4mm}
\end{table}

\subsection{Measuring NIDS Performance: Key Metrics}
To evaluate the effectiveness of Network Intrusion Detection Systems (NIDS), we delve into essential metrics derived from the confusion matrix. This matrix succinctly captures the system's accuracy in classifying network traffic. In our analysis, we focus on the following critical parameters:
\begin{enumerate}
    \item \textbf{Accuracy:} Measures the overall correctness of the NIDS, reflecting the proportion of correct predictions (both attacks and normal traffic) out of all predictions.
    \begin{equation*}
        \text{Accuracy} = \frac{\text{(TP + TN)}}{\text{(TP + TN + FP + FN)}}
    \end{equation*}
    \item \textbf{Recall (also known as Detection Rate or Sensitivity):} Measures the NIDS's ability to correctly identify actual attacks. It answers the question: How many of the actual attacks were correctly detected?
    \begin{equation*}
        \text{Recall} = \frac{\text{TP}}{\text{TP + FN}}
    \end{equation*}
    \item \textbf{Precision:} Assesses the proportion of predicted attacks that were indeed correct. It focuses on how many of the predicted attacks were true positives.
    \begin{equation*}
        \text{Precision} = \frac{\text{TP}}{\text{TP + FP}}
    \end{equation*}
    \vspace{-2mm}
    \item \textbf{F1-Score:} The F1-score harmoniously combines precision and recall into a single metric, providing a balanced view of the NIDS's performance.
    \begin{equation*}
        \text{F1-Score} = \frac{\text{2 * (Precision * Recall)}}{\text{Precision + Recall}}
    \end{equation*}
\end{enumerate}
While overall accuracy is crucial for assessing Network Intrusion Detection Systems (NIDS), its limitations become evident in imbalanced datasets. In such scenarios, we prioritize recall to minimize missed attacks, as even a single undetected threat can have significant consequences. Conversely, high precision is essential to avoid false alarms and maintain operational efficiency. Ultimately, the F1-score, which balances attack detection with false alarm mitigation, provides a nuanced understanding of NIDS efficacy for robust network security.

\section{EXPERIMENTAL RESULTS AND ANALYSIS}
Our experimental investigation, conducted using Python and TensorFlow on a standard personal computer, explored the viability of Lightweight Convolutional Neural Networks (CNNs) for intrusion detection in comparison to traditional Machine Learning (ML) methods. Leveraging the versatile UNSW-NB15 dataset for multi-class classification, we constructed a customized CNN architecture within TensorFlow. This architecture allowed us to rigorously evaluate its performance against established techniques. By directly comparing their effectiveness and resource requirements, our study provides valuable insights for deploying CNNs in real-world network security scenarios. The following are the hyperparameters of our model during the training phase: Binary classification( loss='binary\_crossentropy', optimizer='adam', Learning rate= 0.001, Input shape= (1, 20)) and Multi-classification(loss='categorical\_crossentropy', optimizer='adam', Learning rate= 0.01, Input shape= (1, 20))

\begin{figure}[t!]
\vspace{4mm}
	\centering
 \includegraphics[width=0.9\linewidth]{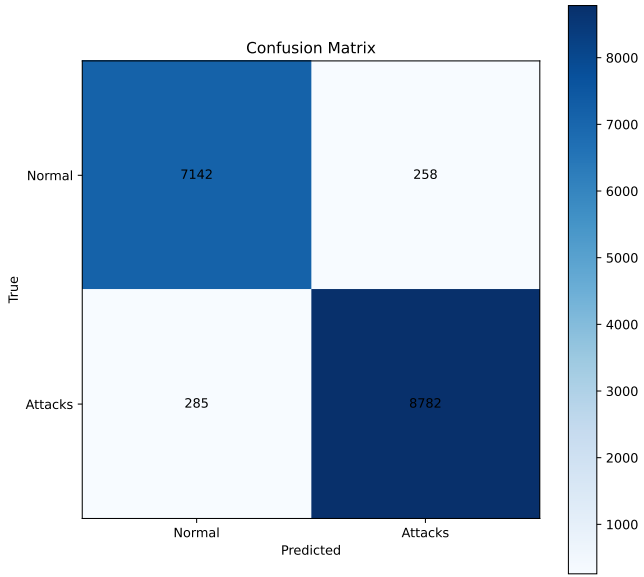}
	\caption{CNN-BiLSTM confusion matrix for binary classification of attacks.}
	\label{matrix}
 \vspace{-6mm}
\end{figure}

The performance of our proposed model in classifying attacks on the UNSW-NB15 dataset is illustrated in the confusion matrix depicted in Figure \ref{matrix}. The X-axis represents the model's predicted attack categories for the testing data, while the Y-axis displays the actual attack categories based on the ground truth labels. Our model successfully identified 7,142 true positives and 8,782 true negatives, accurately distinguishing between normal and attack instances. Nonetheless, it also produced 258 false negatives (missed attacks) and 285 false positives (incorrectly identified attacks). Despite achieving an overall accuracy of 97\%, as indicated in Table \ref{eval}, we are dedicated to exploring strategies to mitigate the remaining false positives and negatives.

\begin{table*}[t!]
\centering
\vspace{4mm}
\renewcommand{\arraystretch}{1.1}
\caption{Comparing Binary Classification Performance: Proposed Model vs. Existing Models on UNSW-NB15 Dataset}
\label{eval}
\begin{tabular}{l c c c c c c c }
\toprule
 & \textbf{Accuracy} & \textbf{Recall} & \textbf{Precision} & \textbf{F1-Score} & \textbf{time to train (s)} & \textbf{time to predict (s)} & \textbf{total time (s)} \\ \bottomrule
\textbf{Logistic Regression} & 92.80\% & 92.80\% & 92.83\% & 92.80\% & 1.2 & 0.0 & 1.2 \\ 
\textbf{KNN} & 95.04\% & 95.04\% & 95.09\% & 95.05\% & 0.0 & 15.3 & 15.3 \\ 
\textbf{Decision Tree} & 96.56\% & 96.56\% & 96.56\% & 96.56\% & 1.0 & 0.0 & 1.0 \\ 
\textbf{Gradient Boosting Classifier} & 95.85\% & 95.85\% & 95.86\% & 95.85\% & 27.4 & 0.0 & 27.4 \\ 
\textbf{MLP} & 96.34\% & 99.99\% & 55.05\% & 70.60\% & 21.3 & 21.9 & 43.2 \\  
\textbf{LSTM}  & 96.49\% & 96.49\% & 96.49\% & 96.49\% & 70.9 & 22.4 & 93.3 \\ 
\textbf{CNN-BiLSTM \cite{IWCMC-JOUHARI}} & 97.28\% & 96.44\% & 98.59\% & 97.43\% & 122.5 & 3.8 & 126.3 \\
\textbf{Proposed model} & 97.90\% & 97.90\% & 97.91\% & 97.90\% & 53.5 & 1.1 & 54.6 \\\bottomrule
\end{tabular}
\vspace{-5mm}
\end{table*}

Table \ref{eval} illustrates the performance evaluation of our CNN-BiLSTM model using feature selection in comparison to traditional ML models found in existing literature. The evaluation included metrics such as accuracy, recall, precision, and F1-score. While precision and recall are commonly utilized for imbalanced datasets, we focused on precision to assess our model's ability to prevent overfitting. By combining these metrics into the F1-score, we gain a holistic view of our model's performance. Additionally, we took into account training and inference times to showcase the suitability of our proposed model for resource-constrained IoT devices. The results in the table indicate that our model surpasses other models in terms of accuracy, achieving 97.90\% compared to the top-performing CNN-BiLSTM \cite{IWCMC-JOUHARI} at 97.28\%. This indicates that our model is more adept at detecting attacks on the UNSW-NB15 dataset. Furthermore, our model achieves a high precision at 97.90\%, ensuring precise attack classification. Given our focus on resource-constrained IoT devices in smart cities, we also evaluated the model's training and inference times. In the context of on-device processing, the training duration of 53.5 seconds may appear extensive. However, it is important to note that training typically takes place offline in practical settings, thereby not adversely affecting the performance of our model. Moreover, the inference time of 1.1 seconds stands out as notably quicker compared to other deep learning models, even though it is not specifically optimized for real-time processing. In contrast to the CNN-BiLSTM model proposed in \cite{IWCMC-JOUHARI}. (2019), our model achieves a reduced prediction time by employing feature selection techniques that decrease the number of features inputted into the model. This reduction in computational requirements enhances the suitability of our model for IoT devices with limited resources.

\begin{figure}[t!]
	\centering
	\includegraphics[width=0.98\linewidth]{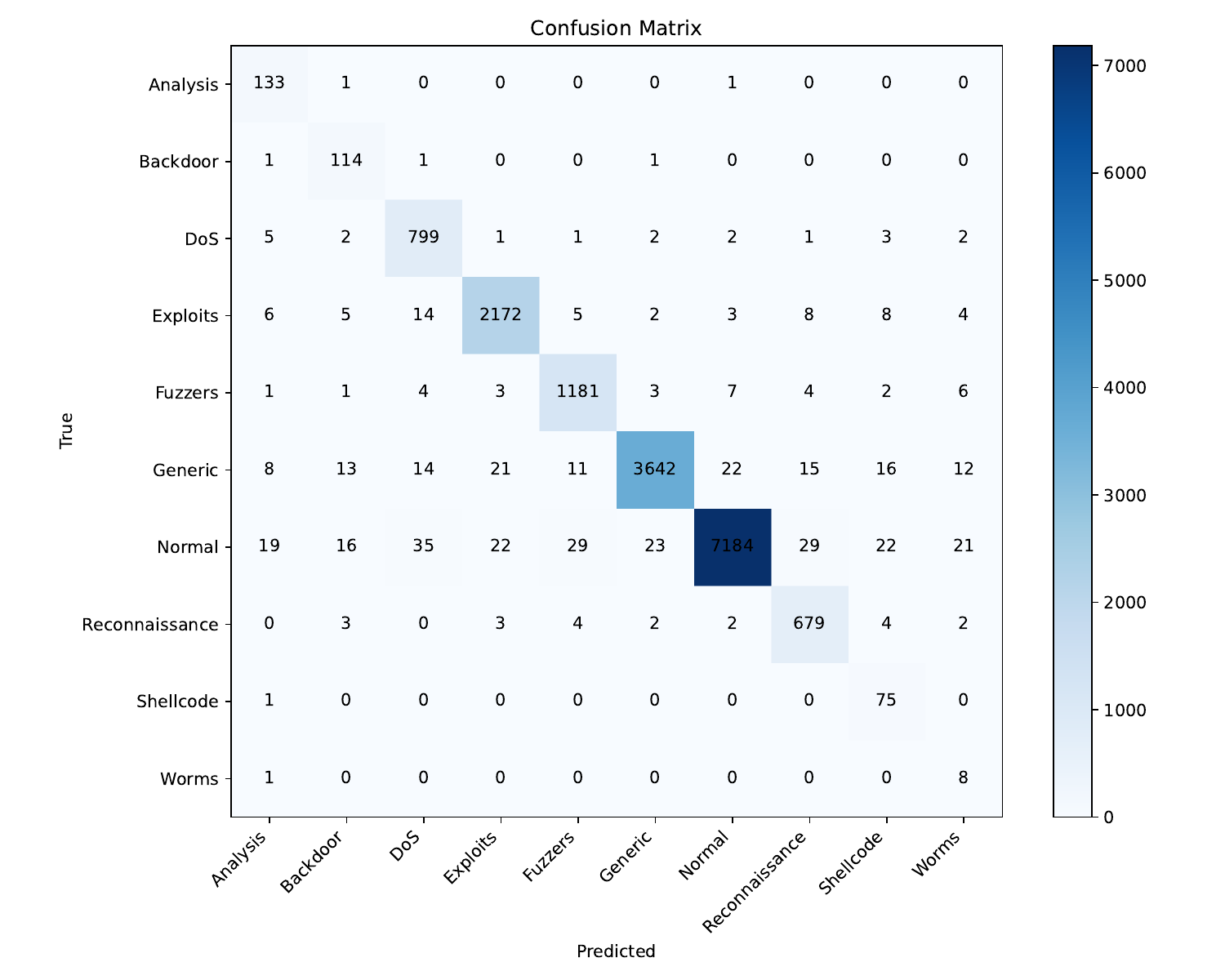}
	\caption{CNN-BiLSTM confusion matrix for multiclassification of attacks.}
	\label{matrix_multi}
 \vspace{-4mm}
\end{figure}

\begin{table*}[t]
    \centering
    \vspace{3mm}
    \renewcommand{\arraystretch}{1.1}
    \caption{Comparing Multi-Classification Performance: Proposed Model vs. Existing Models on UNSW-NB15 Dataset}
    \label{tab:multiclass}
    \begin{tabular}{lcccccc}
    \toprule
 & \textbf{Accuracy} & \textbf{Precision} & \textbf{Recall} & \textbf{F1-Score} & \textbf{Time to Train (s)} & \textbf{Time to Predict (s)} \\ \bottomrule
\textbf{Logistic} & 91.83\% & 94.31\% & 92.74\% & 93.53\% & 2.20 & 0.10 \\
\textbf{KNN} & 93.06\% & 95.06\% & 93.80\% & 94.43\% & 0.30 & 10.10 \\
\textbf{Decision Tree} & 84.10\% & 89.89\% & 86.28\% & 88.09\% & 0.20 & 0.10 \\
\textbf{GBC} & 75.47\% & 85.16\% & 79.11\% & 82.09\% & 55.90 & 0.10 \\
\textbf{MLP} & 96.74\% & 97.45\% & 97.00\% & 97.22\% & 122.80 & 0.10 \\
\textbf{LSTM} & 96.11\% & 97.02\% & 96.43\% & 96.72\% & 90.80 & 15.10 \\
\textbf{CNN-BiLSTM \cite{IWCMC-JOUHARI}} & 96.91\% & 97.54\% & 97.13\% & 97.33\% & 1220.20 & 3.20 \\
\textbf{Proposed Model} & 97.09\% & 97.09\%	& 97.62\%	 & 97.27\% & 440.10 & 2.10 \\ \bottomrule
\end{tabular}
\vspace{-5mm}
\end{table*}

The confusion matrix depicted in Figure \ref{matrix_multi} provides a comprehensive overview of the performance of our CNN-BiLSTM multiclassification model on the UNSW-NB15 dataset. This matrix enables us to assess the classification accuracy for each attack class individually. The dataset encompasses a total of 10 attack classes, as indicated by the x and y labels. The x-axis represents the ground truth or the true labels, while the y-axis displays the predicted classes generated by our model. The diagonal line, extending from the top-left to the bottom-right, signifies the model's proficiency in accurately classifying each attack type. The intensity of the squares in the matrix corresponds to the accuracy level, with darker squares indicating higher accuracy for the respective class. However, it is important to note that the reliability of square colors may be compromised for minority classes such as "Analysis," "Shellcode," and "Worms" due to the limited number of samples available in the test data. To address this issue, our model incorporates a weighted loss function, which prioritizes achieving improved accuracy on these minority classes. The accurate identification of 8 of 9 worms and the successful categorization of 75 out of 76 Shellcode attacks, with only a single false negative, clearly demonstrates this fact. In contrast, the normal traffic class exhibits the darkest square on the matrix, owing to its exceptional precision and ample number of samples. In essence, the confusion matrix provides a visual depiction of the model's effectiveness, where a darker diagonal indicates superior overall performance.

Table \ref{tab:multiclass} presents the outcomes of the multiclassification process conducted on our proposed model, in comparison to the traditional model. To achieve these results, we made certain modifications to our model while maintaining the same architecture, as opposed to the binary classification. These adjustments primarily involved the hyperparameters of the model, such as the number of epochs, batch size, and learning rate. Consequently, these alterations had an impact on both the training and prediction durations. The multiclassification results were comparatively lower than those attained in the binary classification due to the intricate nature of the task, which entails classifying 10 different attacks using our proposed model. Nevertheless, the obtained outcomes serve as evidence that our model surpasses other models in terms of accuracy, precision, recall, and F1-score, achieving respective values of 97.09\%, 97.09\%, 97.62\%, and 97.27\%. Despite the fact that the proposed model necessitates more time during the training phase due to adjustments in the number of epochs and batch size, this process is always conducted offline before deploying the model on resource-constrained IoT devices. Once deployed, these devices function as Intrusion Detection Systems (IDS), and even with the increased training time and resource consumption, our model still fulfills the requirements of these devices in terms of resource usage. This is due to the relatively lower time and resource requirements for prediction. In contrast, employing feature selection alongside CNN-BiLSTM resulted in a decrease in inference time from 3.20 to 2.10 when compared to the findings of \cite{IWCMC-JOUHARI}. This signifies that the utilization of feature selection has improved the applicability of the CNN-BiLSTM based IDS for IoT devices that have limited resources.

\vspace{-2mm}
\section{Conclusion}
In this study, our primary focus was on developing a lightweight IDS model tailored to the computational constraints of IoT devices. This crucial aspect is often neglected in the creation of complex machine-learning models aimed at enhancing IDS performance. Consequently, existing models in the literature were largely unsuitable for deployment on resource-limited devices. Hence, our proposed model was designed to enhance accuracy while accounting for model complexity and the inference time. To achieve this, we merged a lightweight CNN-BiLSTM with $\chi^2$ feature selection to create an accurate and less complex IDS. The utilization of feature selection techniques decreased the number of features inputted into the model, subsequently reducing the model's complexity by cutting down on the inference time. This was demonstrated through the comparison of our model's prediction time of 2.10 with that of \cite{IWCMC-JOUHARI} at 3.20, which utilized all the features of the UNSW-NB15 dataset. Through our experiments, we conducted binary and multiclass classification of attacks using the UNSW-NB15 dataset. This study marks the inception of a new research avenue that we intend to pursue further to enhance accuracy and delve deeper into the complexity and computational demands of the model. 

\vspace{-2mm}
\bibliographystyle{ieeetr}
\bibliography{library}

\begin{thebibliography}{10}

\bibitem{10122600}
M.~Jouhari, N.~Saeed, M.-S. Alouini, and E.~M. Amhoud, ``A survey on scalable lorawan for massive iot: Recent advances, potentials, and challenges,'' {\em IEEE Communications Surveys \& Tutorials}, vol.~25, no.~3, pp.~1841--1876, 2023.

\bibitem{s19081935}
S.~Li, H.~Song, and M.~Iqbal, ``Privacy and security for resource-constrained iot devices and networks: Research challenges and opportunities,'' {\em Sensors}, vol.~19, no.~8, 2019.

\bibitem{10279198}
M.~Jouhari, K.~Ibrahimi, J.~B. Othman, and E.~M. Amhoud, ``Deep reinforcement learning-based energy efficiency optimization for flying lora gateways,'' in {\em ICC 2023 - IEEE International Conference on Communications}, pp.~6157--6162, 2023.

\bibitem{9685598}
H.~Benaddi, M.~Jouhari, K.~Ibrahimi, and A.~Benslimane, ``Securing iot transactions against double-spending attacks based on signaling game approach,'' in {\em 2021 IEEE Global Communications Conference (GLOBECOM)}, pp.~1--6, 2021.

\bibitem{jouhari:hal-04432030}
M.~Jouhari, K.~Ibrahimi, H.~Tembine, M.~Benattou, and J.~Ben~Othman, ``{Signaling game approach to improve the MAC protocol in the underwater wireless sensor networks},'' {\em {International Journal of Communication Systems}}, vol.~32, June 2019.

\bibitem{Omarov2023}
B.~Omarov, O.~Auelbekov, A.~Suliman, and A.~Zhaxanova, ``Cnn-bilstm hybrid model for network anomaly detection in internet of things,'' {\em International Journal of Advanced Computer Science and Applications}, vol.~14, no.~3, 2023.

\bibitem{10180926}
P.~R. Agbedanu, R.~Musabe, J.~Rwigema, I.~Gatare, and Y.~Pavlidis, ``Ipca-samknn: A novel network ids for resource constrained devices,'' in {\em 2022 2nd International Seminar on Machine Learning, Optimization, and Data Science (ISMODE)}, pp.~540--545, 2022.

\bibitem{9768133}
E.~Gyamfi and A.~D. Jurcut, ``Novel online network intrusion detection system for industrial iot based on oi-svdd and as-elm,'' {\em IEEE Internet of Things Journal}, vol.~10, no.~5, pp.~3827--3839, 2023.

\bibitem{10.1007/s00500-023-09452-7}
M.~Al-Ambusaidi, Z.~Yinjun, Y.~Muhammad, and A.~Yahya, ``Ml-ids: an efficient ml-enabled intrusion detection system for securing iot networks and applications,'' {\em Soft Comput.}, vol.~28, p.~1765–1784, dec 2023.

\bibitem{tatsunami2022sequencer}
Y.~Tatsunami and M.~Taki, ``Sequencer: Deep {LSTM} for image classification,'' in {\em Advances in Neural Information Processing Systems} (A.~H. Oh, A.~Agarwal, D.~Belgrave, and K.~Cho, eds.), 2022.

\bibitem{IWCMC-JOUHARI}
M.~Jouhari and M.~Guizani, ``Lightweight cnn-bilstm based intrusion detection systems for resource-constrained iot devices,'' in {\em 2024 International Wireless Communications and Mobile Computing (IWCMC)}, 2024.

\end{thebibliography}
\end{document}